\documentclass[12pt]{article}
\usepackage[dvips]{color}
\usepackage{epsfig}
\usepackage{amsmath}
\usepackage{graphicx}

\begin{document}
\title{Hydrodynamics in black brane with hyperscaling violation metric background}
\author{{J. Sadeghi $^{a}$\thanks{Email: pouriya@ipm.ir} and A. Asadi
$^{b}$\thanks{Email: ali.asadi89@stu.umz.ac.ir}}\\\\
{$^{a}$ \emph{Department of Physics, Islamic Azad University -
Ayatollah Amoli Branch,}}\\
{\emph{P.O.Box 678, Amol, Iran}}\\
{$^{b}$ \emph{Physics Department, University of Mazandaran },}\\{
\emph{P.O.Box 47416-95447, Babolsar, Iran}}} \maketitle
\begin{abstract}
In this paper we consider a metric with hyperscaling violation on
black brane background. In this background we calculate the ratio of
shear viscosity to entropy density with hydrodynamics information.
The calculation of this quantity lead us to constraint $\theta$ as
$3\leq\theta<4$, and $\theta\leq0$. In that case we show that the
quantity of $\frac{\eta}{s}$ not dependent to hyperscaling violation
parameter $\theta.$ Our results about ratio of shear viscosity to
entropy density in direct of $QCD$ point of view  agree with other
works in literature as $1/4\pi$.
\noindent\\
\\
{\bf PACS number:} 11.25.Uv; 12.36.Mh; 51.20.+d \\
{\bf Keywords:} Hyperscaling violation; Shear viscosity; Diffusion;
Hydrodynamic.
\end{abstract}
\section{Introduction}
One of the important subject in modern theoretical physics is the
understanding the dynamics of strongly coupled quantum field
theories. Also there is a large class of real world physics systems
which is described by the perturbation theory as $QCD$ system. One
of the interesting case is the state of matter discovered in heavy
ion collisions, the Quark Gluon Plasma $(QGP)$, that is known as a
nearly ideal fluid. Then it is logical to look at this state of
matter as a
hydrodynamics theory.\\
Hydrodynamics is an effective theory to describe the macroscopic
dynamics at large distances and time scales. The conserved
quantities such as energy-momentum tensor $T^{\mu\nu}$ are
considered to survive in such large distances and time scales.
Unlike the familiar effective field theories the hydrodynamics is
normally formulated in the language of equations of motion instead
of an action principle, because there is dissipation in thermal
media. The description in this section is basically following the
review papers \cite{Banerjee,Takeuchi,Son}. The hydrodynamic
equation by conservation of energy-momentum tensor is given by,
\begin{eqnarray}\label{con}
\partial_\mu T^{\mu\nu}=0.
\end{eqnarray}
The conformal hydrodynamic in $d$ dimensional space-time at local
rest frame is described by $d$ independent variable: temperature
$T(x)$ and the $d$-velocity vector field $u^\mu(x)$ , which is
satisfied by $u_\mu u^\mu = -1$. So we express  $T^{\mu\nu}$ in
terms of $T(x)$ and $u^\mu(x)$. the hydrodynamic quantities in local
rest frame are varying slowly over space-time, therefor their
derivatives are very small. So, we can express the energy-momentum
tensor in power of spatial derivative of local quantities. At the
zeroth order (ideal fluid), $T^{\mu\nu}$ is given by,
\begin{eqnarray}\label{ideal}
T^{\mu\nu}=(\epsilon+P)u^\mu u^\nu+Pg^{\mu\nu},
\end{eqnarray}
where $\epsilon$ is energy density, $P$ is the pressure and
$g^{\mu\nu}$ is the metric tensor background space-time. At the next
order in derivative expansion fluid energy-momentum tensor is given
by,
\begin{eqnarray}\label{vis}
T^{\mu\nu}&=&(\epsilon+P)u^\mu u^\nu+Pg^{\mu\nu}-\sigma^{\mu\nu},\nonumber\\
\sigma^{\mu\nu}&=&P^{\mu\alpha}P^{\nu\beta}\left\{\eta\left(\nabla_\alpha
u_\beta+\nabla_\beta u_\alpha-\frac{2}{d-1}g_{\alpha\beta}\nabla
u\right)+\zeta g_{\alpha\beta}\nabla u\right\},\nonumber\\
P^{\mu\nu}&=&g^{\mu\nu}+u^\mu u^\nu,
\end{eqnarray}
where $\sigma^{\mu\nu}$ is proportional to derivative of $T(x)$ and
$u^\mu(x)$, and it is the dissipative part of energy-momentum
tensor, also $\sigma^{\mu\nu}$ is a symmetric tensor. The
coefficients $\eta$ and $\zeta$ are called shear and bulk viscosity.
If the system contains a conserved current, there is an additional
hydrodynamic equation related to the current conservation $j^\mu$
with $\partial_\mu j^\mu=0$. The constitutive equation is given as,
\begin{eqnarray}\label{ideal}
\emph{\textbf{j}}=\rho \emph{\textbf{u}}-\mathcal{D}\nabla j^t,
\end{eqnarray}
where $j^t=\rho$ and $\mathcal{D}$ are charge density and the
diffusion constant respectively. In the fluid rest frame,
$\emph{\textbf{j}}=-\mathcal{D}\nabla\rho$, which is Fick's law of
diffusion.\\
The hydrodynamics behavior of a system is determined by transport
coefficients, shear viscosity, bulk viscosity and etc. As we
mentioned the $QGP$ produced at Relativistic Heavy Ion Collision
$(RHIC)$ treat as nearly perfect fluid or a viscus fluid with very
small shear viscosity. With a low ratio of shear viscosity to
entropy density is very hard to describe with normal methods. The
temperature of $QGP$ produced at $RHIC$ is almost $170 MeV$ that, is
so close to confinement temperature of $QCD$. So, in high
temperature they are not in the weakly coupled area of $QCD$. In
fact they are close to transition temperature of $QGP$ in
non-perturbative regime of $QCD$. So, the normal calculation of
perturbative gauge theory is not suitable for describing $QGP$. One
important implement for understanding the dynamics of strongly
coupled is $AdS/CFT$ correspondence
\cite{Maldacena1,Gubster,Witten}. In general the $AdS/CFT$
correspondence relates between two system, it means that the
interacting quantum field theory one hand and string theory in a
curved background on
the other hand.\\
In Ref.\cite{Starinets} the authors by using the $AdS/CFT$
correspondence calculated the shear viscosity of the
finite-temperature $N=4$ supersymmetric Yang-Mills theory in the
large N in non-extremal black 3-brane. Also in Refs.[1-7] are used
of the $AdS/CFT$ technique for calculating of hydrodynamics
quantities. Computing of the second order derivative of transport
coefficient is shown in
Refs.\cite{Loganayagam,Bhattacharyya,Banerjee2}. In Ref.
\cite{Kovtun} the authors describe the infrared  behavior of
theories whose dual gravity that contain a black brane with non-zero
Hawking temperature. They argue that the infrared behavior of these
theories is survive by nothing other than hydrodynamics. Also, they
obtained a general relation for transport coefficients (diffusion
rate, shear viscosity)in terms of the components of the metric for a
large class of metrics. In Ref.\cite{Kovtun2} has been shown that
the ratio of shear viscosity to entropy density has a lower bound as
$\frac{\eta}{s}\geq \frac{1}{4\pi}$. Also this quantity obtained for
theories such as $D_3$-brane, $M_2-brane$, $M_5-brane$, $D_p-brane$
\cite{Son,Policastro,Kovtun}. But such calculation did not attend
for theories with hyperscaling violation metric background. The
general review on holographic hydrodynamics is provided in
\cite{Banerjee}. All the above information give us motivation to
calculate the diffusion constant and ratio of shear viscosity to
entropy density. So, in this paper, we review of the hyperscaling
violation metric background, and also consider a black brane
solution of this metric. Next we study the hydrodynamics of this
metric background by using the diffusion constant and calculate
ratio of shear viscosity to entropy density.
\section{Review hyperscaling violation metric background}
Here we are going to introduce the $AdS$ metric background as,
\begin{eqnarray}\label{AdS}
ds^2=\frac{R^2}{z^2}\left(-dt^2+dx_i^2+dz^2\right),
\end{eqnarray}
where $i=1..3$ , $R$ is the $AdS$ scale and $z$ is holographic
coordinate. This metric is invariant under a dilation of all
coordinate $z \rightarrow \lambda z$ and $x_\mu \rightarrow \lambda
x_\mu$. One generalization of AdS gravity theory and their conformal
field theory dual, is to consider metrics which have reduced
symmetries compared to anti de-Sitter space, and these theories
display Lifshitz scaling symmetry with following metric
\cite{Dong,Narayan},
\begin{eqnarray}\label{Lifshitz}
ds^2=-\frac{1}{r^{2\tau}}dt^2+\frac{1}{r^2}\left(dr^2+dx^2_i\right),
\end{eqnarray}
which $\tau$ is dynamical exponent ($\tau=1$ gives the AdS metric).
This metric background is invariant under following scaling,
\begin{equation}\label{scale3}
t\rightarrow\lambda^\tau t,\qquad x_i\rightarrow\lambda x_i,\qquad
 r\rightarrow\lambda r.
\end{equation}
These metrics are produced in effective gravity theories with a
negative cosmological constant,  and with abelian gauge fields in
the bulk. By containing an abelian gauge field and scalar dilation,
the full class of these metrics are \cite{Charmousis},
\begin{equation}\label{metric4}
ds^2=r^{-2{{(d-\theta)}/{d}}}\left(r^{-2(\tau-1)}dt^2+dr^2+dx^2_i\right),
\end{equation}
where $\theta$ is hyperscaling violation exponent. This metric is
not invariant under scale transformation (\ref{scale3}), but
transform as,
\begin{equation}\label{scale5}
ds=\lambda^{\theta/d}ds,
\end{equation}
To consider a physically sensible dual field theories from gravity
side, we should product the null energy condition as $T_{\mu\nu}
n^{\mu}n^{\nu}\geq 0$, where the null vectors satisfy
$n^{\mu}n_{\mu}=0$ \cite{Dong,Narayan}. From these condition we
obtain ,
\begin{eqnarray}\label{NEC1}
&(d-\theta)(d(\tau-1)-\theta)\geq0,\nonumber\\
&(\tau-1)(d+\tau-\theta)\geq0.
\end{eqnarray}
Aspect of holography for hyperscaling violation metric presented in
Ref. \cite{Dong}. Starting from metric (\ref{metric4}) with
hyperscaling violation (by fixing $\tau=1$), the black brane
solution become \cite{Dong,Sadeghi},
\begin{eqnarray}\label{metric5}
ds^2&=&\left(\frac{r_F}{r}\right)^{2\theta/d}\left\{\left(\frac{r^2}{R^2}\right)\left(-fdt^2+dx_i^2\right)
+\left(\frac{R^2}{r^2}\right)f^{-1}dr^2+R^2d\Omega_5^2\right\},\nonumber\\
f(r)&=&1-\left(\frac{r_H}{r}\right)^{4-\theta},
\end{eqnarray}
where $i=1..3$. $r_H$ is the radius of horizon and $r_F$ is a
dynamical scale which come from dynamical analysis. this is
responsible for restoring the canonical dimensions in presence of
hyperscaling violation. We note that by setting $\theta=0$, metric
background (\ref{metric5}) reduce to non-extremal $D3$-brane
geometry in the near horizon.
\section{Hydrodynamics and ratio of shear viscosity to entropy density}
As we know the  hydrodynamics is an effective theory for the
describing  the macroscopic dynamics at large distances and it is a
proper method for explain of matter discovered in heavy ion
collision $(QGP)$. In the hydrodynamics we have some important
quantities such as shear viscosity, which plays important role in
physics of early universe. One of most important problems of $QGP$
is  shear viscosity, such  quantity  play important role in strongly
coupled thermal gauge theories  which is  achieved by the $AdS/CFT$
correspondence. There are several approach to obtain the ratio of
shear viscosity to entropy density. The most known approach for the
such quantity is Kubo formula . But in this paper, in order to
obtain the ratio of shear viscosity to entropy density we use
diffusion constant for the corresponding background.  Now we are
going to calculate  the ratio of shear viscosity to entropy density
for metric background (\ref{metric5}) as Ref.\cite{Kovtun}. Here we
note that,  for general metric background we have following
expression, $ds^2=g_{00}(r)dt^2+g_{rr}(r)dr^2+g_{xx}(r)dx_i^2$, so,
the  diffusion constant can be obtained by following relation
\cite{Kovtun},
\begin{eqnarray}\label{diffusion1}
\mathcal{D}=\frac{\sqrt{-g(r_H)}}{g_{xx}(r_H)g_{eff}^2(r_H)\sqrt{-g_{00}(r_H)g_{rr}(r_H)}}\int_{r_H}^\infty
dr\frac{-g_{00}(r)g_{rr}(r)g_{eff}^2}{\sqrt{-g(r)}},
\end{eqnarray}
where $g_{eff}$ is an effective gauge coupling which is coming from
the action of the gauge field dual to the conserved current and, can
be a function of the radial coordinate $r$. In generally we can say
that for the obtaining the $\mathcal{D}$, we have to give
perturbation to the corresponding background metric. If we give
perturbation to diagonal component $g_{eff}=$const and for metric
background (\ref{metric5}) one can obtain $\mathcal{D}$ as,
\begin{eqnarray}\label{diffusion2}
\mathcal{D}=r_H^{1-\theta/3}R^2\int_{r_H}^\infty dr
r^{\theta/3-3}=\frac{1}{2-\theta/3}\frac{R^2}{r_H}\qquad
and,\qquad\theta<6.
\end{eqnarray}
As we know in equation (\ref{NEC1}) we got $\tau=1$ correspond to
metric background (\ref{metric5}). In that case we have two
conditions in $d=3$,
\begin{eqnarray}\label{NEC2}
\theta\leq0\qquad and,\qquad\theta\geq3.
\end{eqnarray}
These two conditions and the finite form of integral
(\ref{diffusion2}) and equation (\ref{NEC2}) lead us to pick up
$\theta$ as following,
\begin{eqnarray}\label{NEC3}
3\leq\theta<6\qquad and,\qquad\theta\leq0.
\end{eqnarray}
Here also note that in case of $\theta=0$ on can achieve the
$\mathcal{D}$ as a $D3$-brane system \cite{Kovtun}. In second step
we are going to calculate the shear mode of the stress-energy
tensor. In that case we have effectively $g_{eff}^2=g_{xx}$
\cite{Son,Kovtun}. So, we have
\begin{eqnarray}\label{diffusion3}
\mathcal{D}=r_H^{3-\theta}R^2\int_{r_H}^\infty dr
r^{\theta-5}=\frac{1}{4-\theta}\frac{R^2}{r_H}\qquad
and,\qquad\theta<4,
\end{eqnarray}
where by considering of the conditions of (\ref{NEC2}), we have,
$3\leq\theta<4$, and $\theta\leq0$. Now we are going to investigate
the ratio of shear viscosity to entropy density, we use
$\eta/s=T\mathcal{D}$ ($T$ is the Hawking temperature)\cite{Son,Kovtun}.\\
 The Hawking temperature of metric
background (\ref{metric5}) can be obtained by surface gravity as,
\begin{eqnarray}\label{surface1}
\kappa^2=-\frac{1}{2}(\nabla^\mu \chi^\nu)(\nabla_\mu
\chi_\nu),\qquad \beta=\frac{1}{T}=\frac{2\pi}{\kappa},
\end{eqnarray}
where $\kappa$ is the surface gravity and $\chi$ is the killing
vector field. Since the metric background (\ref{metric5}) is
independent of time,  so $\partial/\partial t$ is a killing vector
for this metric. Thus we can obtain the surface gravity and
temperature  as,
\begin{eqnarray}\label{surface2}
\kappa=\frac{1}{2}\sqrt{-g^{tt}g^{rr}(g_{tt,r})^2}=\frac{4-\theta}{2}\frac{r_H}{R^2},\qquad
T=\frac{4-\theta}{4\pi}\frac{r_H}{R^2}
\end{eqnarray}
where $\theta=0$ give the result of $D3$-brane geometry. So by using
the equations (\ref{diffusion2}), (\ref{diffusion3}) and
(\ref{surface2}) one can obtain the diffusion constant as
$\mathcal{D}=\frac{4-\theta}{2-\theta/3}\frac{1}{4\pi T}$ and
$\mathcal{D}=\frac{1}{4\pi T}$. We note here in case of shear mode
the diffusion constant not dependent to hyperscaling violation. Also
from relations of (\ref{diffusion3}) and (\ref{surface2}) we obtain
the ratio of shear viscosity to entropy density,
\begin{eqnarray}\label{vis.ent}
\frac{\eta}{s}=\frac{1}{4\pi}
\end{eqnarray}
In generally, the ratio of shear viscosity to entropy density not
dependance to hyperscaling violation in case of this paper. Also, we
can say that the $\theta$ will be constraint with
$3\leq\theta<4$, and $\theta\leq0$.\\
\section{Conclusion}
In this work, we considered hyperscaling violation metric
background. In this background we calculated the   ratio of shear
viscosity to entropy density with hydrodynamics information. The
calculation of this quantity lead us to constraint $\theta$ as
$3\leq\theta<4$, and $\theta\leq0$. In that case we have shown that
such quantity with the first order calculation of hydrodynamics not
dependent to hyperscaling violation parameter $\theta.$ Our results
about ratio of shear viscosity to entropy density complectly agree
with other works as \cite{Son,Kovtun}. The advantage of this
approach for the calculation of $\mathcal{D}$ and $\frac{\eta}{s}$
is much simpler than of AdS/CFT and Kubo's approaches.

\end{document}